\documentclass[twocolumn,twocolappendix]{aastex631}
\usepackage{placeins}
\usepackage{amsmath} 
\usepackage{multirow}
\usepackage{subfigure}
\usepackage{enumerate}
\usepackage{ragged2e} 

\newcolumntype{P}[1]{>{\Centering\arraybackslash}p{#1}} 

\newcommand{\new}[1]{\textcolor{black}{{#1}}}

\begin{document}

\title{Wave generation via oscillatory reconnection at a three-dimensional magnetic null point}

\author[0000-0002-5082-1398]{Luiz A. C. A. Schiavo}
\affiliation{Northumbria University, Newcastle upon Tyne, NE1 8ST, UK}

\author[0000-0002-5915-697X]{Gert J. J. Botha}
\affiliation{Northumbria University, Newcastle upon Tyne, NE1 8ST, UK}

\author[0000-0002-7863-624X]{James A. McLaughlin}
\affiliation{Northumbria University, Newcastle upon Tyne, NE1 8ST, UK}



\begin{abstract}
This work conducts a three-dimensional (3D), nonlinear magnetohydrodynamic (MHD) simulation to investigate wave generating, time-dependent reconnection around a magnetic null point. A non-periodic perturbation (in the $xz$-plane) triggers oscillatory reconnection (OR) at the 3D null, resulting in \new{a self-sustained {oscillation with a constant period} $P$. We investigate} the response of the system using three distinct wave proxies (compressible parallel, compressible transverse and incompressible parallel) as well as Spectral Proper Orthogonal Decomposition for decoupling and analyzing the resultant MHD wave behavior. We find that OR generates a slow magnetoacoustic wave of period $P$ that propagates outwards in all directions along the spine and fan plane of the 3D null point. We also find the generation of a propagating Alfv\'en wave of period $P$, exclusively along the $y$-axis in the fan plane, i.e. in the direction perpendicular to the spine motion. These findings provide new insights into waves generated from a 3D null point and their implications for coronal seismology.
\end{abstract}



\section{Introduction} \label{sec:intro}
Magnetic reconnection is a key mechanism responsible for converting magnetic energy into other forms such as thermal and kinetic energy, while simultaneously driving particle acceleration and modifying magnetic topology \citep{2022LRSP...19....1P,BROWNING2024100049}. This process underlies many solar phenomena, including coronal mass ejections \citep{2012LRSP....9....3W} and the energy release associated with solar flares \citep{2017LRSP...14....2B}. At smaller scales, observations of chromospheric anemone jets provide evidence that local reconnection events in the lower solar atmosphere play a role in heating the chromosphere and corona \citep{Shibata2007}. Magnetic null points are omnipresent from the photosphere to the upper corona \citep{regnier2008} and function as natural sites where reconnection occurs.

Among the reconnection regimes, oscillatory reconnection (OR) represents a particular form of time-dependent reconnection, characterised by periodic changes in magnetic connectivity. The phenomenon was first reported by \cite{Craig1991}, who analyzed the relaxation of a two-dimensional (2D) X-point magnetic field configuration. OR is characterized by periodic behavior that arises naturally from the system’s internal relaxation dynamics rather than from any external periodic driver.
This self-sustaining oscillation allows OR to produce quasi-periodic outputs even when the initial perturbations are aperiodic \citep{McLaughlin2009,McLaughlin2012a}.

\begin{figure*}[htb]
	\centering
     \includegraphics[trim = 0 0 0 0, clip, width=0.99\textwidth]{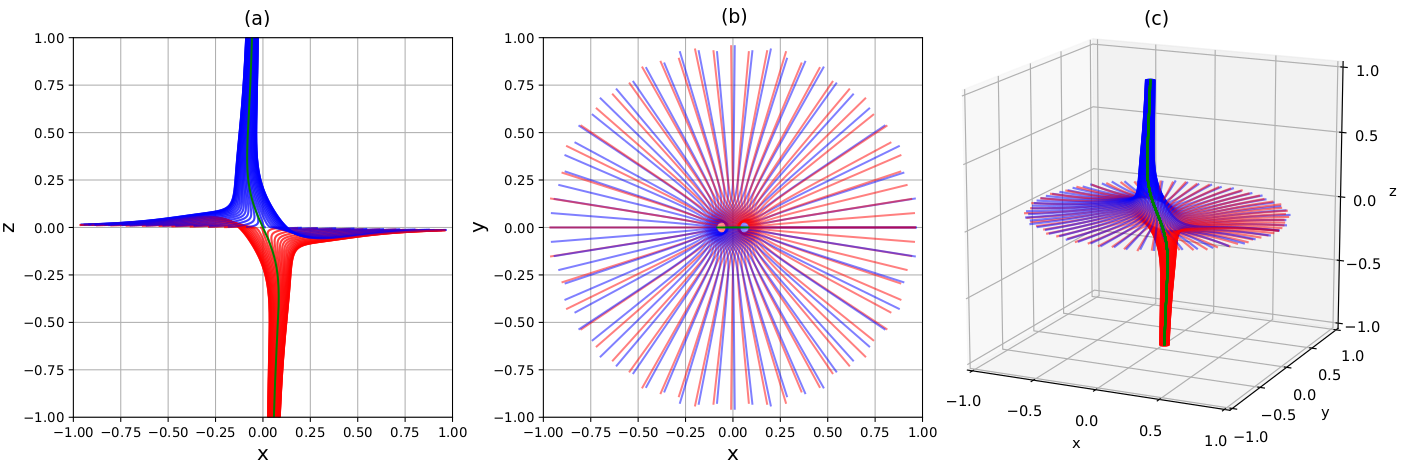}
	\caption{Traced magnetic field lines for the initial condition, panels show different views of the initial condition: (a) a $xz-$view, (b) a $xy-$view, (c) 3D view. The green line represents the null point spine, while the blue and red lines represent the fan plane traced from the upper and lower boundaries, respectively.}
	\label{fig:initial-field}
\end{figure*}

The connection between OR and wave processes becomes particularly intriguing in regions close to magnetic nulls, where magnetohydrodynamic (MHD) mode conversion can occur. This interaction occurs when Alfvén and sound speeds are comparable, a condition often satisfied near null points. 2D numerical experiments have demonstrated how fast-wave fronts wrap around nulls, leading to energy focusing and complex field behavior \citep{mclaughlin2006,nakariakov_quasi-periodic_2006,mclaughlin2011}.

The accumulation of fast-wave energy near nulls intensifies wave amplitudes, frequently producing steepening, shocks, and current density enhancements \citep{nakariakov_quasi-periodic_2006}. Quantitative studies by \cite{Tarr2017} revealed the energy partitioning during such interactions, showing that approximately 70\% of incident wave energy converts to slow magnetoacoustic waves, 7\% to fast magnetoacoustic waves, with the remaining 23\% persisting at the null for eventual dissipation. Recent advances in both simulations and observations have strengthened these findings.  \cite{mondal2024} simulated external perturbations interacting with a 2D coronal null, observing current sheet formation and subsequent plasmoid dynamics that generate fast magnetoacoustic waves, while \cite{kumar2024} provided the first observational evidence of fast-to-slow mode conversion at a 3D null point within a pseudostreamer configuration, bridging the gap between theoretical predictions and solar observations.

Moving to three dimensions (3D) introduces additional complexity. In 3D systems, magnetic reconnection can develop in current layers that either contain null points or not, and field lines evolve through continuous slippage rather than discrete reconnection of pairs \citep{priest_nature_2003}. At 3D nulls, reconnection may occur through different modes such as spine-fan, torsional spine, or torsional fan reconnection \citep{priest_three-dimensional_2009}.


Despite these advances, significant gaps remain in our understanding of time dependent 3D magnetic reconnection. Previous 3D studies have focused on analytical frameworks \citep{priest_three-dimensional_2009}, steady-state models \citep{wyper_torsional_2010,wyper_torsional_2011}, or on tearing-mode simulations \citep{wyper_dynamic_2014,huang_turbulent_2016}. The first simulation of OR in 3D was reported by \citet{Thurgood2017}, who perturbed a 3D null point using a driver perpendicular to the spine, triggering an oscillatory response. \citet{schiavo2025OR} extended these analyses to longer evolution times, demonstrating that the oscillation period remains constant and independent of the driver amplitude. Follow-up simulations by \citet{sabri_plasma_2021,sabri_propagation_2022} used an Alfvénic driver in similar setups. Building upon these studies, the present work investigates the wave patterns produced throughout the OR cycle.

The paper is organized as follows: \S\ref{sec:numerical-model} details our numerical approach; \S\ref{sec:results} contains the results, which are divided into the analysis of the OR signature and field line motion in \S\ref{sec:OR}, a study of the waves along the spine in \S\ref{sec:waves-spine}, and fan plane in \S\ref{sec:waves-fan}; with conclusions in \S\ref{sec:conclusions}.


\section{Methodology and diagnostics} \label{sec:numerical-model}
\subsection{MHD simulation}
The 3D resistive MHD equations are solved using the Lare3D numerical code \citep{Arber2001}. This code employs a Lagrangian predictor–corrector step followed by an Eulerian remap, ensuring conservation of mass, momentum, and energy. In the present study, the response of a 3D magnetic null point was simulated, that is perturbed by a localized spherical magnetoacoustic pulse, which initiates OR.

\begin{figure}[htb]
\includegraphics[ width=0.94\columnwidth]{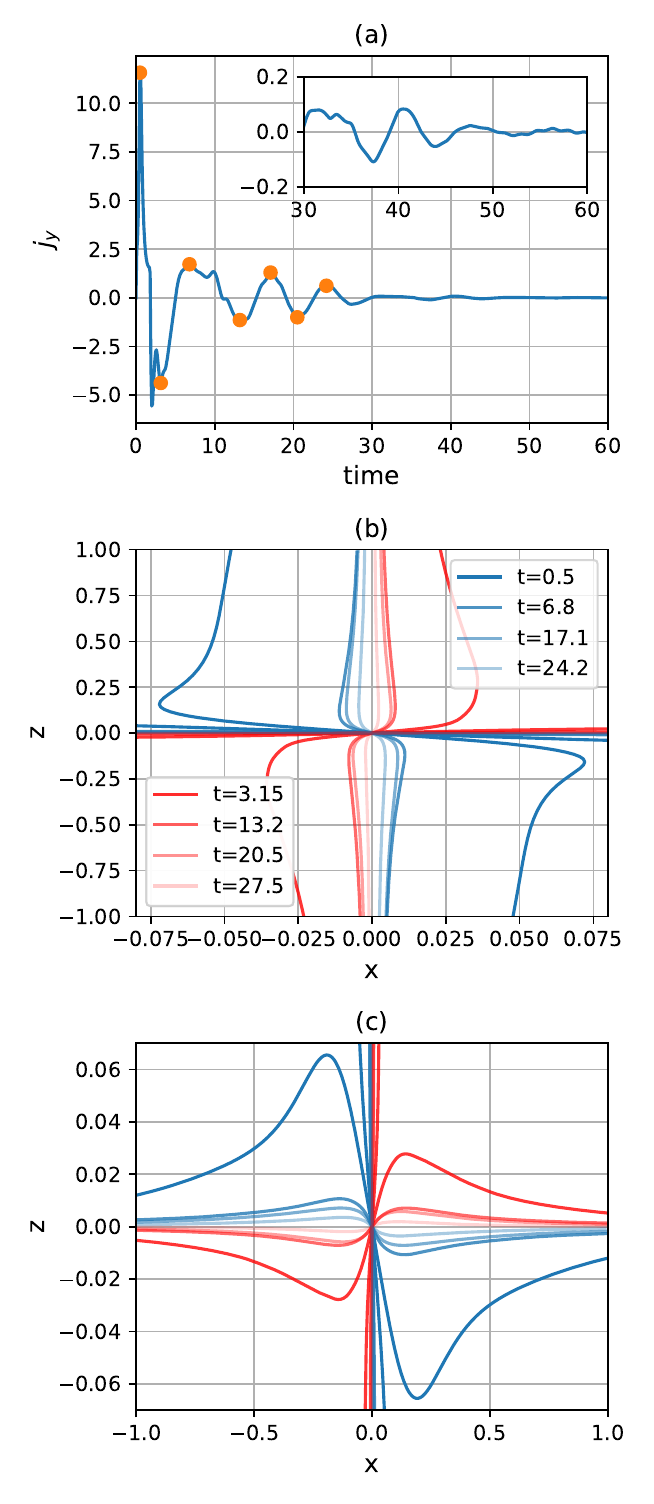}
\caption{(a) Current density oscillations at the null point, $j_y(0,0,0,t)$, showing a signature of oscillatory reconnection, where the blue curve represents the simulated $j_y(0,0,0,t)$ and orange dots indicate the simulation times displayed in panels (b)-(c) and Figures \ref{fig:slices}, \ref{fig:slices-fan} and \ref{fig:fan-MHD-alfven}. (b) Spine evolution during reconnection cycles, with blue and red curves corresponding to the $j_y(0,0,0,t)$ orange dots in panel (a). (c) Fan plane evolution at $y=0$, using the same color scheme from panel (b) to denote the oscillation phases. Field lines from panels (b)-(c) were traced from the null point.}
	\label{fig:jy}
\end{figure}

The governing equations, written in their dimensionless Lagrangian form, are expressed as
\begin{eqnarray}
\frac{D\rho}{Dt} &=& -\rho \nabla \cdot \mathbf{v}, \label{eq:mass}\\
\frac{D\mathbf{v}}{Dt} &=& \frac{1}{\rho}(\nabla\times\mathbf{B})\times\mathbf{B}
    -\frac{1}{\rho}\nabla p + \frac{1}{\rho}\mathbf{F}_{\mathrm{visc}}, \label{eq:momentum}\\
\frac{De}{Dt} &=& -\frac{p}{\rho}\nabla\cdot\mathbf{v}
    +\frac{\eta}{\rho}|\mathbf{j}|^{2}
    +\frac{1}{\rho}Q_{\mathrm{visc}}, \label{eq:energy}\\
\frac{D\mathbf{B}}{Dt} &=& (\mathbf{B}\cdot\nabla)\mathbf{v}
    -\mathbf{B}(\nabla\cdot\mathbf{v})
    -\nabla\times(\eta\nabla\times\mathbf{B}), \label{eq:induction}\\
p &=& \rho e(\gamma-1), \label{eq:state}
\end{eqnarray}
where $D/Dt$ denotes the material derivative, $\rho$ the mass density, $\mathbf{v}$ the velocity field, and $\mathbf{B}$ the magnetic field. The electric current density is given by $\mathbf{j}=\nabla\times\mathbf{B}$, $p$ is the thermal pressure, $e$ is the specific internal energy, $\eta$ is the magnetic diffusivity, and $\gamma=5/3$ is the adiabatic index. \new{The viscous force $\mathbf{F}_{\mathrm{visc}}$ and the associated heating term $Q_{\mathrm{visc}}$ provide numerical stabilization and represent artificial viscous dissipation 
\citep{Caramana1998}.}

The system is nondimensionalized by characteristic scales of length $L_0$, magnetic field strength $B_0$, and mass density $\rho_0$, with $\mu_0$ denoting the magnetic permeability of free space. These define the derived units $v_0 = B_0/\sqrt{\mu_0\rho_0}$, $t_0 = L_0/v_0$, $j_0 = B_0/(\mu_0 L_0)$, $p_0 = B_0^2/\mu_0$, and $e_0 = v_0^2$. \new{A detailed explanation for the boundary conditions and setup for this simulation are described in \citet{schiavo2025OR}.}


\subsection{Initial magnetic field configuration}\label{sec:2.2}
The computational domain contains a single 3D magnetic null point located at the origin of a Cartesian coordinate system. The equilibrium field corresponds to a linear, potential, proper null configuration \citep{1996PhPl....3..759P}. In this geometry, the fan surface coincides with the $z=0$ plane, while the spine axis aligns approximately with the $z$-direction. This configuration is identical in form to that used by \citet{Thurgood2017} and \citet{schiavo2025OR}, and  the present work focuses on the associated and generated wave dynamics.

The initial magnetic field is decomposed into a background equilibrium and a perturbation field,
\begin{equation}
\mathbf{B} = \overline{\mathbf{B}} + \mathbf{B}' , \label{eq:B_total}
\end{equation}
with
\begin{equation}
\overline{\mathbf{B}} = (x,\, y,\,-2z), \qquad
\mathbf{B}' = \nabla\times\mathbf{A}', \label{eq:B_components}
\end{equation}
and
\begin{equation}
\mathbf{A}' = \psi\, \exp\!\left[-\frac{x^{2}+y^{2}+z^{2}}{2\sigma^{2}}\right] \hat{\mathbf{y}} .
\label{eq:A_potential}
\end{equation}
Here, overbars denote equilibrium quantities and primes indicate perturbations. \new{The decomposition is exact and the perturbations are nonlinear.} The equilibrium plasma parameters are uniform, with $\overline{\rho}=1$, $\overline{\mathbf{v}}=0$, and $\overline{p}=0.005$, corresponding to a plasma beta of $\beta=0.01$ at a distance of unity from the null. \new{The resistivity is constant and set to be $\eta=10^{-3}$. }

The perturbation amplitude and spatial extent are controlled by the constants $\psi$ and $\sigma$, respectively. Unless otherwise stated, $\sigma=0.21$ and $\psi=0.05$ were used, following \new{\citet{Thurgood2017}. 
The perturbation is the same as that in \citet{Thurgood2017} and \citet{schiavo2025OR}, and is selected to induce a slight bend  in the spine relative to the fan, in the $xz-$plane, aiming to trigger oscillatory spine-fan reconnection.}
Simulations are evolved for a total time of $60\,t_0$. Figure~\ref{fig:initial-field} shows the resulting magnetic topology based on Eqs.~(\ref{eq:B_total})--(\ref{eq:A_potential}); the perturbation produces a noticeable bending of the spine field lines away from perfect alignment with the $z$-axis.

\begin{figure}[htb]
\includegraphics[ trim =0 0 0 0,clip, width=0.99\columnwidth]{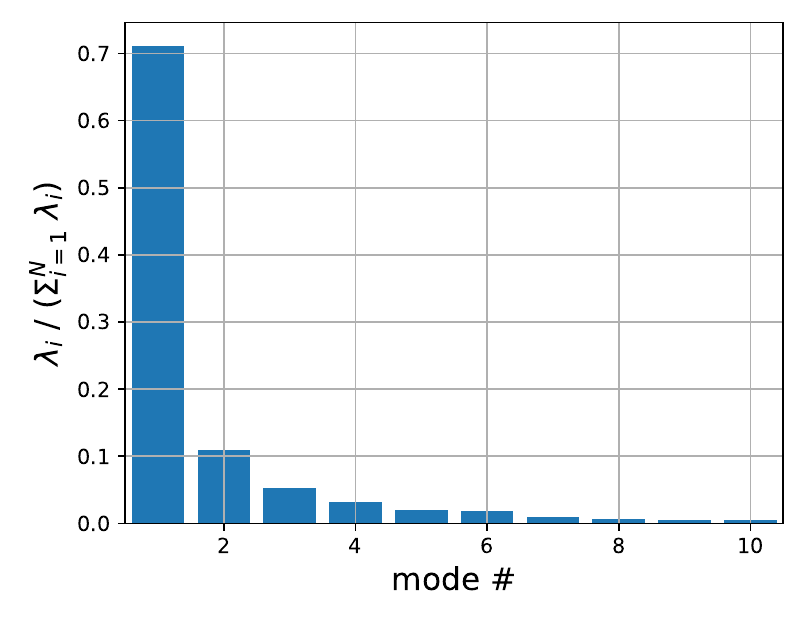}
\caption{SPOD energy spectra, where mode energy, $\lambda_i$, is normalized by the total perturbation energy, $\sum_i^N \lambda_i$.}
	\label{fig:pod-energy}
\end{figure}

\subsection{MHD wave proxies in 3D} \label{sec:wave-split}

In a uniform plasma, Alfvén waves are incompressible and are parallel to vorticity, while magnetoacoustic waves (fast and slow) are compressible waves that in general do have a parallel component of displacement but do not propagate parallel to vorticity \citep{goossens_mixed_2019}. In a non-uniform plasma, MHD waves are coupled and can have mixed properties. In low-$\beta$ plasmas, such as the solar corona, slow magnetoacoustic waves manifest as longitudinal, sound-like disturbances that are aligned with the magnetic field. Fast waves, by contrast, propagate most rapidly in the direction transverse to the magnetic field. 

These wave characteristics are quantified using three key identifiers. Following the formalism of \cite{Raboonik2022} \new{we define three MHD wave-proxies:}
\begin{align}
\xi_A &= (\nabla \times \mathbf{v}) \cdot \mathbf{e}_{\|}, \label{eq:alfven} \\
\xi_{\|} &= \nabla \cdot (v_{\|} \mathbf{e}_{\|}), \label{eq:parallel} \\
\xi_{\perp} &= \nabla \cdot \mathbf{v} - \xi_{\|} = \nabla \cdot(\mathbf{v}-v_\|\mathbf{e}_\|) . \label{eq:perp}
\end{align}
Here, $\mathbf{e}_{\|} = \mathbf{B}/|\mathbf{B}|$ is the unit vector parallel to the magnetic field $\mathbf{B}$. The variable $v_{\|}=\mathbf{v}\cdot\mathbf{e}_{\|}$ denotes the field-aligned velocity component. These quantities serve as proxies for three distinct wave components and were applied previously on wave identification \citep{Raboonik2022,enerhaug2024,enerhaug2025}. \new{These wave-proxies have a physical meaning since they represent: an incompressible parallel component ($\xi_A$, Eq.\ \ref{eq:alfven}), a compressible parallel component ($\xi_{\|}$, Eq.\ \ref{eq:parallel}), and a compressible transverse component ($\xi_{\perp}$, Eq.\ \ref{eq:perp}). These identifiers are not exact representations of the three fundamental MHD wave modes, but they encapsulate the essential features distinguishing wave behavior, and hence are of use in mode identification.}

\begin{figure}[htb]
\includegraphics[ trim =13 13 11 10,clip, width=0.99\columnwidth]{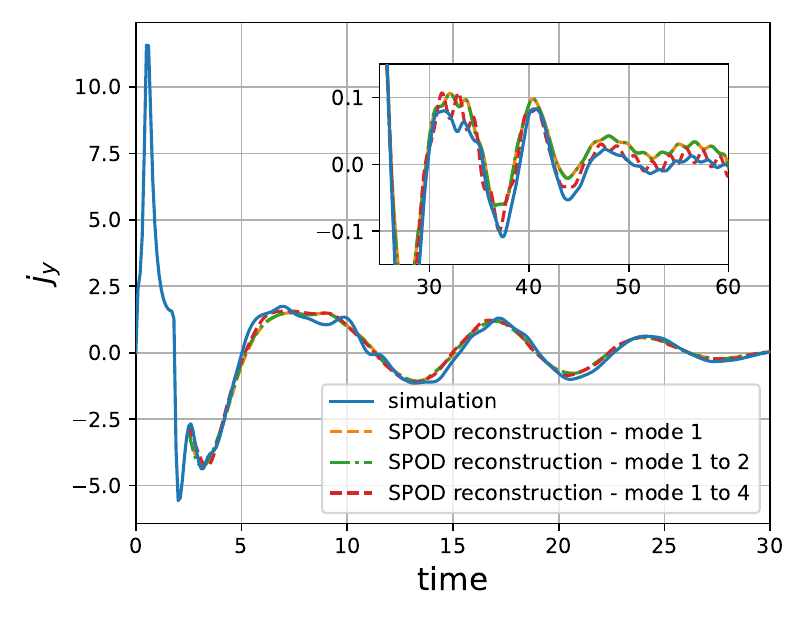}
\caption{Comparison between simulation and SPOD for the current density measured at the null point.}
	\label{fig:pod-jy}
\end{figure}

\subsection{Spectral Proper Orthogonal Decomposition (SPOD)} \label{sec:POD}

Proper orthogonal decomposition (POD) is a powerful dimensionality reduction technique that can extract coherent structures from high-dimensional systems. Originally applied by \citet{Lumley1967} for analyzing turbulent fluid flows, this method has demonstrated remarkable success in solar physics applications. It has been used for analysis of large-scale magnetic structures (sunspots and pores) and has enabled the first identification of multiple high-order eigenmodes in photospheric observations \citep{2021RSPTA.37900181A, 2022ApJ...927..201A, jafarzadeh_wave_2025}. Here SPOD is applied from \citet{Sieber2016}, which is an evolution of POD that provides a more statistically robust way to isolate coherent structures within specific frequency bands, compared to POD.

SPOD is especially valuable for analyzing complex, non-stationary systems with inhomogeneous structures. The method's core principle involves decomposing a time series of fluctuating fields, typically represented as discrete snapshots, into a hierarchical series of orthogonal spatial modes. These modes are ranked by their associated eigenvalues, which quantitatively represent each mode's contribution to the total variance (or energy content) of the system. The decomposition yields two key components for each mode: the spatial structures (spatial mode) that reveal coherent patterns, and temporal coefficients that describe the mode's evolution.

\begin{figure}[htb]
\includegraphics[trim = 0 0 0 0, clip, width=0.95\columnwidth]{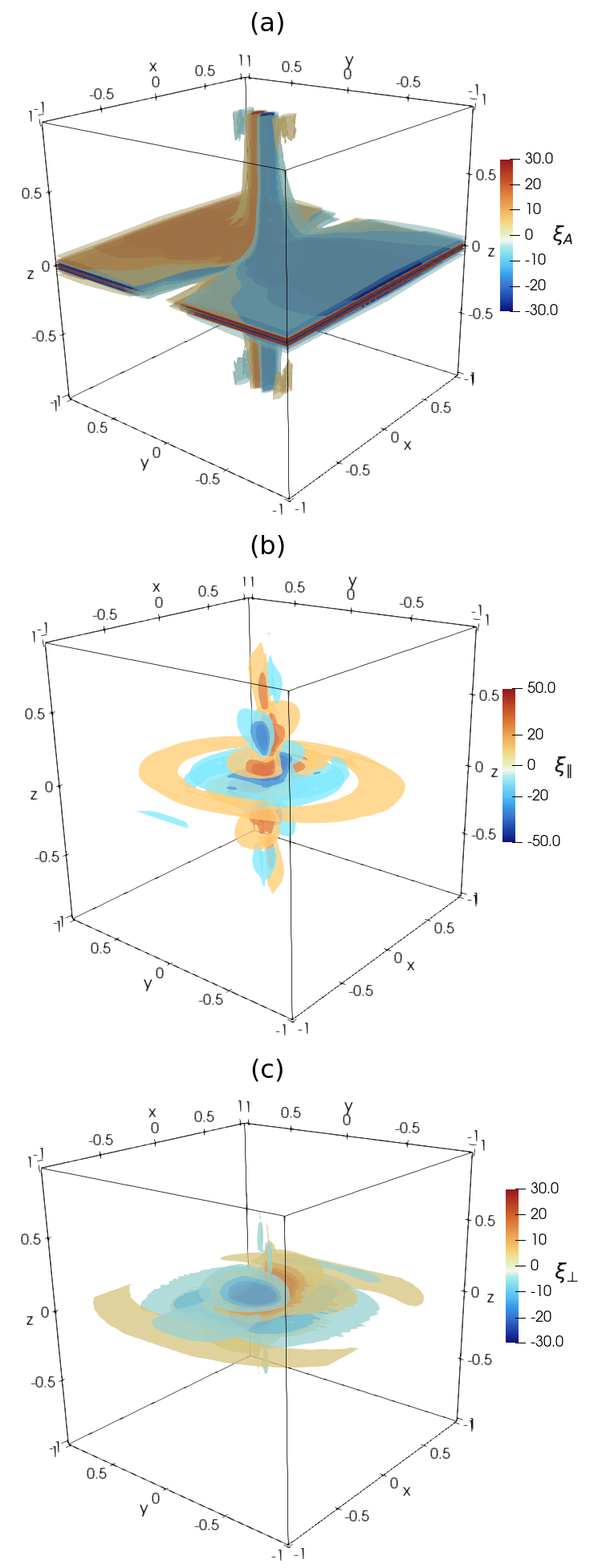}
	\caption{Isosurfaces of SPOD dominant \new{spatial mode $\mathbf{\phi(\mathbf{x})}$} for the MHD wave proxies: (a) $\xi_A$, (b) $\xi_\|$ and (c) $\xi_\bot$.}
	\label{fig:POD-modes3D}
\end{figure}

Crucially, the eigenvalue spectrum provides immediate physical insight: modes with larger eigenvalues capture more significant contributions to the system's overall dynamics. This ranking enables efficient low-dimensional representations by retaining only the most energetically important modes while still capturing the dominant features of the flow or field evolution. The decomposition of a vector of variables, $\mathbf{Q}$, is done as:
\begin{eqnarray}
\mathbf{Q}\left(\mathbf{x},t \right) & = & \left\langle \mathbf{Q} \left(\mathbf{x}\right) \right\rangle 
+ \mathbf{Q}^{\prime\prime}\left(\mathbf{x},t \right) , \\
\mathbf{Q}^{\prime\prime}\left(\mathbf{x},t \right) & = &   \sum_{n=1}^{N} a^{\left( n \right)} \left(t \right) \boldsymbol{\phi}^{\left( n \right)} \left(\mathbf{x} \right)\mbox{ ,} 
\end{eqnarray}
where $\boldsymbol{\phi}^{\left( n \right)}$ represents a set of space-dependent orthonormal modes, $a^{(n)}$ is a time-dependent mode amplitude, $N$ is the number of snapshots, and $n$ identifies the mode index, $\left\langle \, \, \right\rangle$ represents a time average, and the double prime denotes a fluctuation. 
A reconstructed fluctuation field, represented by $\widetilde{( \, \, )}$, can then be approximated by:
\begin{equation}
\widetilde{\mathbf{Q}}^{\prime\prime}\left(\mathbf{x},t \right) \; \approx \; \sum_{n=1}^{M} a^{\left( n \right)} \left(t \right) 
\boldsymbol{\phi}^{\left( n \right)} (\mathbf{x}) \mbox{ ,}
\end{equation}
where $M < N$ is the number of modes used in the reconstruction. Note that the time-averaged fluctuation relates to the initial base state fluctuation as $\mathbf{Q}^{\prime}+\overline{\mathbf{Q}}=\mathbf{Q}^{\prime\prime}+\left\langle \mathbf{Q} \right\rangle $. Using the snapshots method introduced by \citet{Sirovich1987}, the modal basis was constructed using a covariance matrix of the magnetic fluctuation field as:
\begin{equation}
C_{t_{1},t_{2}} \; = \; \frac{1}{N} \int_{\Omega}\mathbf{B}^{\prime\prime}\left(\mathbf{x},t_{1}  \right)   \cdot
\mathbf{B}^{\prime\prime} \left(\mathbf{x},t_{2}  \right) d\Omega \mbox{ .}
\end{equation}
$\Omega$ is the integration domain, which was chosen to be $-1\le x,y,z\le1$. 
Additionally, for computing SPOD a kernel filter was applied to the covariance matrix $C$ \citep{Sieber2016}. In our analysis, we used a Gaussian filter with a bandwidth of 15 snapshots. If we had not applied this filter to $C$, we would have obtained the standard POD method described by \citet{Lumley1967}.

This matrix is symmetric, positive, and semi-definite. Therefore the eigenvalues and eigenvectors are computed using singular value decomposition. The eigenvalue problem $CF = \Lambda F$ is solved, where $\Lambda$ is a diagonal eigenvalue matrix and $F$ represents the eigenvector matrix of the covariance matrix $C$. Thus, the SPOD spatial modes were computed by a linear combination of the snapshots into an orthonormal set of basis functions:
\begin{equation}
\boldsymbol{\phi}^{\left( n \right)} \left(\mathbf{x} \right) \; = \; 
\frac{1}{\lambda_{n}N} \sum_{k=1}^{N} F_{k,n} \mathbf{Q}^{\prime\prime}\left(\mathbf{x},t_{k}  \right) \mbox{ ,}
\end{equation}
where $\lambda_{n}$ is the $n^{\mbox{th}}$ eigenvalue from the diagonal eigenvalue matrix $\Lambda$. Finally, the time-dependent mode amplitude is given by:
\begin{equation}
a^{\left( n \right)} \left(t_{k} \right) \; = \;  \sqrt{N \lambda_{n}} F_{k,n} \mbox{ .}
\end{equation}

\subsection{Tracing wave paths} \label{sec:tracing}

\begin{figure}[htb]
\includegraphics[trim = 0 0 0 0, clip, width=0.98\columnwidth]{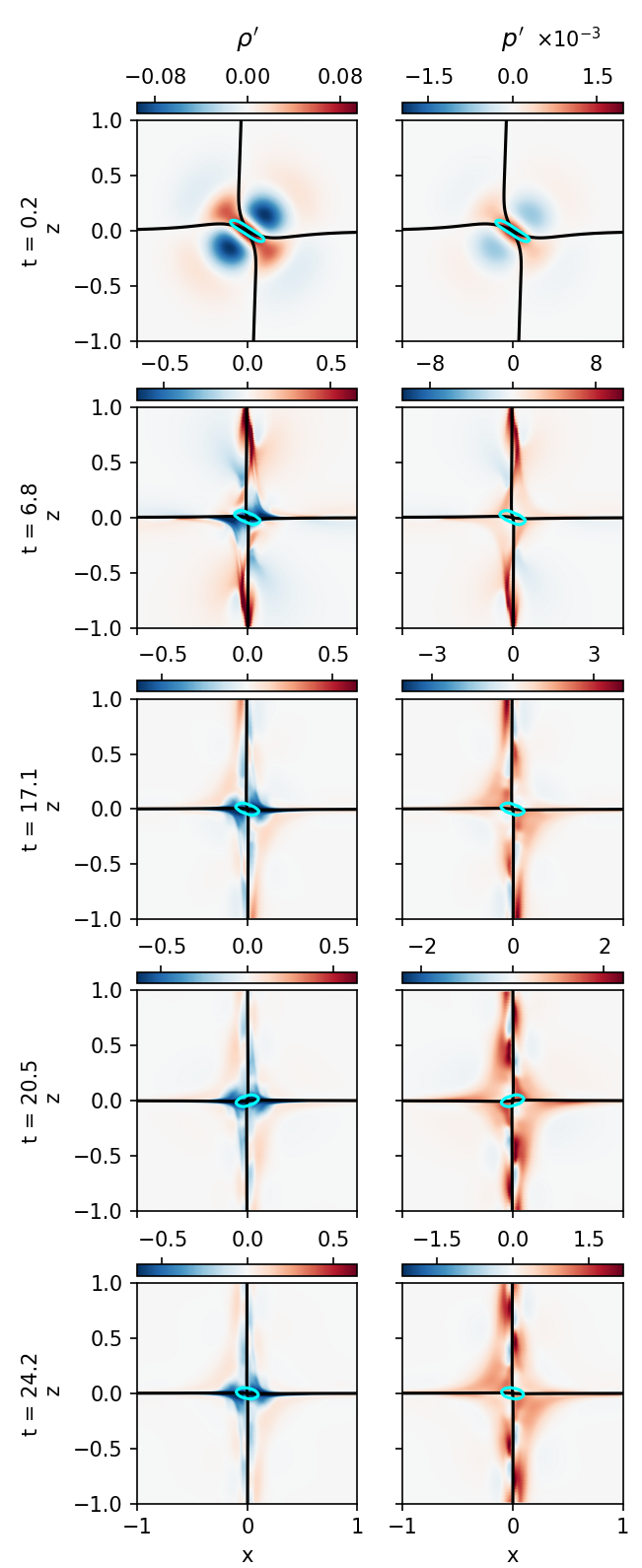}
	\caption{Contour plots in the $y=0$ plane displaying the time evolution of $\rho^\prime$ and $p^\prime$. The cyan lines represent the equipartition layer and the black magnetic field lines indicate the magnetic skeleton. The simulation time instants shown were chosen from the orange dots in Figure \ref{fig:jy}(a).}
	\label{fig:slices}
\end{figure}

To characterize wave propagation time-distance diagrams were created for a specific variable $f(t, S)$ and passive scalar tracers were superimposed on these diagrams. Passive scalar tracers provide insight into the path of a particular wave traveling at a defined reference speed $v_{\text{reference}}$.

The trajectories of the tracers are calculated using a fourth-order Runge-Kutta integration, described by the following equation:\new{
\begin{eqnarray}
\frac{dS}{dt} &=& v_{\text{reference}}(t,S) 
\end{eqnarray}
where the variable $S$ can represent the coordinate along the spine, the $x$-axis, the $y$-axis, or any other specific line. The reference speed $v_{\text{reference}}(t,S)$ depends on space and time,  calculated with the local properties of the full solution.}

The reference speed $v_{\text{reference}}$ could be the local sound speed $v_s = \sqrt{\gamma p/\rho}$, the Alfvén speed $v_A = |\mathbf{B}|/\sqrt{\mu_0 \rho}$, or either the fast $\mathbf{c}_f$ or slow $\mathbf{c}_s$ magnetoacoustic velocities. The components of the magnetoacoustic velocities in 3D are defined as follows:
\begin{align}
c_{f,i}^2 &= \alpha + \sqrt{\alpha^2 - v_s^2\frac{B_i^2}{\mu \rho}} \, ,\\
c_{s,i}^2 &= \alpha - \sqrt{\alpha^2 - v_s^2\frac{B_i^2}{\mu \rho}} \, ,
\end{align} 
where $\alpha=(v_s^2+v_A^2)/2$ and $B_i$ is magnetic field vector in the direction $i$ (see \citet{Raboonik2024}).
propagation


\section{Results} \label{sec:results}

\begin{figure*}[t]
\centering
\includegraphics[trim = 0 0 0 0, clip, width=0.99\textwidth]{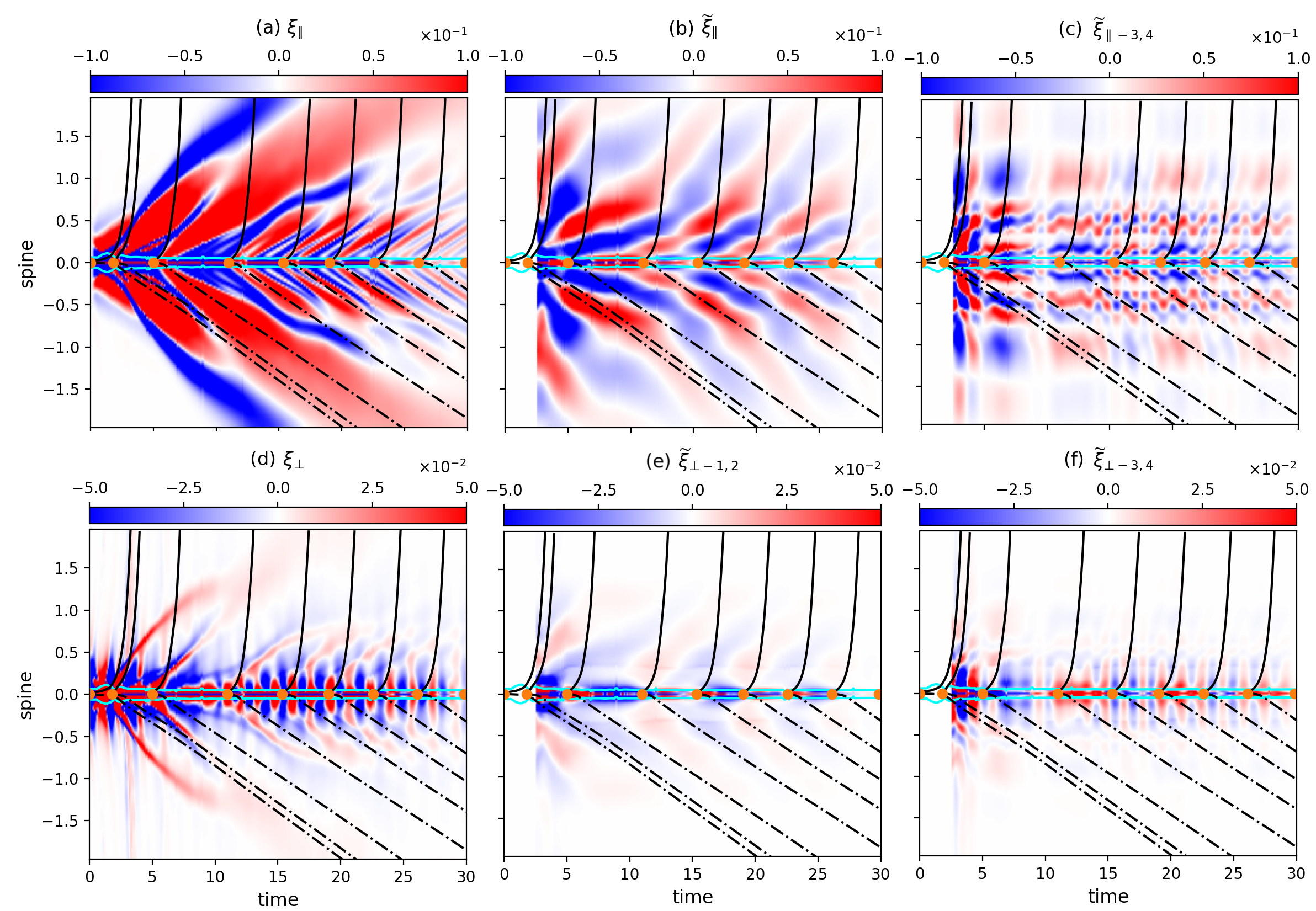}
\caption{Time-distance diagrams for $\xi_\|$ and $\xi_\bot$ and their SPOD reconstruction along the spine. The orange dots mark the roots of $j_y(0,0,0,t)$, while cyan lines indicate the equipartition layer. The black dashed-dotted and black continuous lines represent the slow and fast magnetoacoustic wave trajectories respectively. 
\label{Figure 8 new}
}
\end{figure*}

\subsection{OR signature and field line motion}\label{sec:OR}

The signature of OR is characterized by the oscillatory behavior of the current density at the null point. In this simulation, it was found that the null point remains stationary, located at $x = y = z = 0$, as determined by the null point identification algorithm described in \cite{haynes_trilinear_2007}. Figure \ref{fig:jy}a shows the signature of OR measured \new{over $t = 60$ (with $t = 60$ in nondimensional units)}. When $j_y$ changes its sign, the current sheet is reoriented, as discussed by \cite{McLaughlin2009}.

Early in the simulation (at $t=0.4$) there is a significant overshoot in $j_y(0,0,0,t)$, caused by the initial perturbation. After one complete reconnection cycle, around $t \approx 5$, the influence of the initial perturbation diminishes. At this point, $j_y(0,0,0,t)$ exhibits a more regular oscillatory behavior with a constant frequency and Gaussian decay, as noted by \citet{schiavo2025OR}. The $j_y(0,0,0,t)$ oscillation period $P$ is approximately constant, where \citet{schiavo2025OR} found the period to be $P=(8.1\pm1.1 )t_0$.

\begin{figure*}[htb]
\centering
\includegraphics[trim = 0 0 0 0, clip, width=0.99\textwidth]{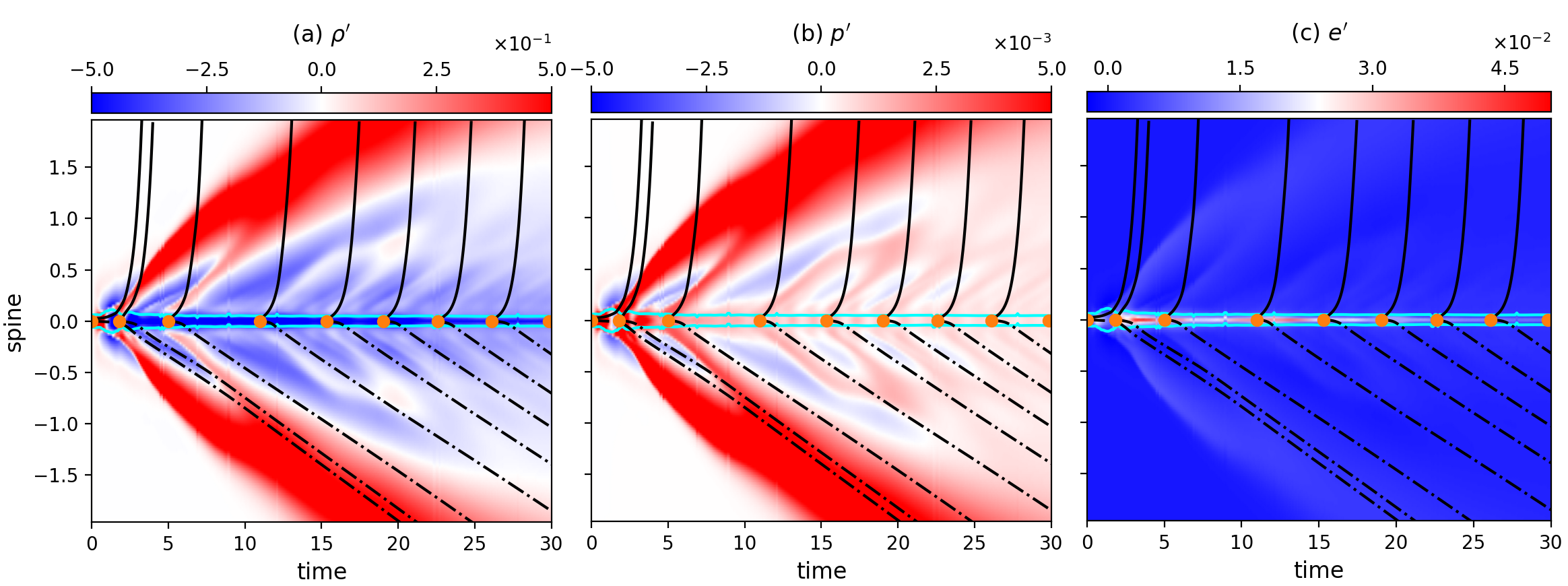}
\caption{Time-distance diagrams showing (a) density, (b) pressure and (c) energy perturbations along the spine. The orange dots mark the roots of $j_y(0,0,0,t)$ in Figure \ref{fig:jy}a, while cyan lines indicate the equipartition layer. The dashed-dotted lines represent trajectories propagating at the local slow magnetoacoustic speed and black continuous lines represent fast magnetoacoustic speed trajectories.}
	\label{fig:st-thermo-spine}
\end{figure*}

Figures \ref{fig:jy}b and~\ref{fig:jy}c illustrate the evolution of the spine and fan during the reconnection cycle. The time instants shown in these figures correspond to the peaks and valleys of $j_y(0,0,0,t)$ marked as orange dots in Figure~\ref{fig:jy}a. 
Peaks are marked with blue lines, while red lines represent valleys in Figures \ref{fig:jy}b and \ref{fig:jy}c.

The current sheet size decreases significantly after the first reconnection cycle, as seen in Figures \ref{fig:jy}b and~\ref{fig:jy}c, as well as the current density amplitude (Figure \ref{fig:jy}a). This reduction is particularly noticeable after $t = 3.2$. Before this time, the initial pulse heavily influences the dynamics (seen in $t = 0.5$ and 3.2). Following the first cycle, the oscillations of $j_y(0,0,0,t)$ exhibit a Gaussian decay, as described by \citet{schiavo2025OR}. These oscillations lead to deformations in the fan plane, which can create propagating disturbances along it. A more detailed analysis of these waves will be presented in the upcoming sections.

SPOD was employed to disentangle perturbation dynamics into coherent modes. The decomposition was computed using 275 simulation snapshots spanning $t=2.5$ to $30$, with the initial time explicitly chosen to avoid contamination from the initial pulse in the SPOD modal basis.

Figure \ref{fig:pod-energy} displays the normalized energy contribution of the first 10 most energetic SPOD modes, where each eigenvalue represents the mode's energy fraction relative to the total fluctuation energy. \new{This result was obtained by integrating over the cube $-1\le x,y,z\le1$ that surrounds the null point.} The histogram reveals three key findings: firstly, mode 1 dominates the system, accounting for approximately 73\% of the total fluctuation energy that corresponds to the OR dynamics. Secondly, the first four modes collectively account for approximately 90\% of the system's fluctuations. Thirdly, higher modes show rapidly decreasing energy contributions, suggesting their minor role in the overall dynamics.

Figure \ref{fig:pod-jy} compares $j_y(0,0,0,t)$ with the SPOD reconstructions, $\widetilde{j}_y(0,0,0,t)$, using only mode 1, as well as modes 1 and 2, and modes 1 to 4 respectively. The reconstruction using only mode 1 closely matches the oscillation pattern of $j_y(0,0,0,t)$, with minor differences. \new{This indicates that mode 1 is sufficient to describe the $\widetilde{j}_y(0,0,0,t)$  signature at the null.} Including modes 2, 3, and 4 results in only slight improvements in the representation of $\widetilde{j}_y(0,0,0,t)$.

Figure \ref{fig:POD-modes3D} presents isosurface visualizations of the dominant SPOD mode (mode 1) for each MHD wave proxy: $\xi_A$, $\xi_{\|}$ and $\xi_{\perp}$ described by Eqs.\ (\ref{eq:alfven})-(\ref{eq:perp}).

For $\xi_A$ (Figure \ref{fig:POD-modes3D}a), associated with incompressible waves, the perturbation mainly occurs in the fan plane with symmetric distribution about $y=0$. The mode exhibits two non-propagating rolls around the spine, consistent with \citet{schiavo2025OR}, that found these vorticity rolls generated by the spine motion, do not propagate along the $z$ direction.
The $\xi_{\|}$ perturbation (Figure \ref{fig:POD-modes3D}b) demonstrates two characteristic features, circular wave patterns in the fan plane and oscillatory behavior along the spine ($z$-axis). This spatial structure suggests coupling between fan plane and spine dynamics, since the oscillations are present in the same mode.
The $\xi_{\perp}$ mode (Figure \ref{fig:POD-modes3D}c) shows more localized behavior, concentrated near the null point and fan plane. In contrast to the parallel perturbations, $\xi_{\|}$, the transverse perturbations do not extend along the $z$-axis (the spine).

\subsection{Wave propagation around the spine} \label{sec:waves-spine}

\begin{figure}[htb]
\includegraphics[trim = 0 0 0 0, clip, width=0.99\columnwidth]{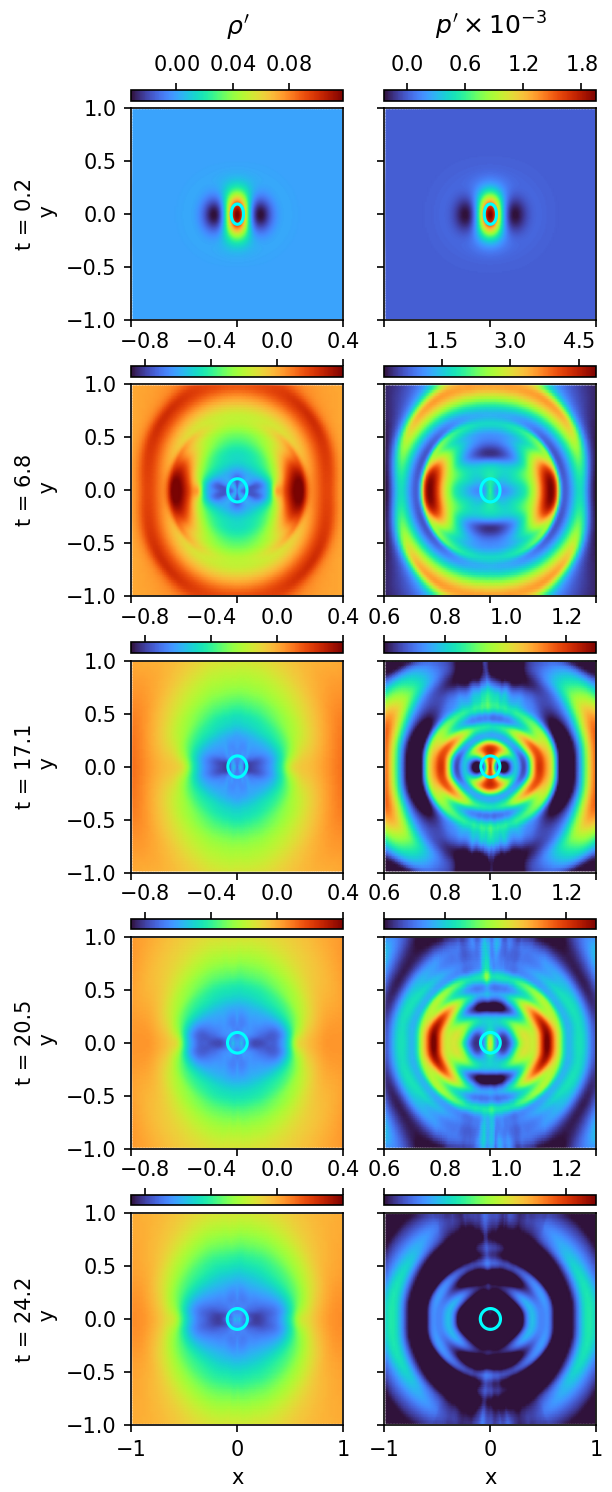}
	\caption{Contour plots in the $z=0$ plane (fan plane) displaying the evolution of $\rho^\prime$ and $p^\prime$. The cyan lines represent the equipartition layer, and the simulation time instants shown were chosen from the orange dots in Figure \ref{fig:jy}a.}
	\label{fig:slices-fan}
\end{figure}

Figure \ref{fig:slices} illustrates contour plots of density and pressure perturbations, defined as $\rho^\prime$ and $p^\prime$. The figure represents evolution along the $y=0$ plane, which is a symmetry plane. 

At $t=0.2$, the initial perturbation generates pressure and density disturbances propagating toward the null point, \new{perturbing the base state}. By $t=6.8$, after the initial pulse has reached the null point and the completion of a single reconnection cycle, there are large perturbations that travel along the spine. At $t=17.1$, during the second reconnection cycle, a distinct wave pattern forms along the spine with a shorter wavelength than that seen at $t=6.8$. It is noteworthy that these pulses are observed in both density and pressure, indicating they are magnetoacoustic waves. The subsequent panels at $t=20.5$ and $t=24.1$ show a similar emission pattern to that observed at $t=17.1$ but with reduced amplitude. Note that the density perturbation amplitude remain similar after the initial transient ($t\le6.8$), while the pressure perturbation reduces by approximately half after each reconnection cycle ($t=6.8$, 17.1 and 24.2).

Additionally, it is significant to highlight that the wave emissions (Figure \ref{fig:slices}) are synchronized with the OR fingerprint from $j_y$ (Figure \ref{fig:jy}a), suggesting that these waves may be generated by the OR, potentially due to reconnection jets reforming in a periodic manner.

\begin{figure*}[t]
\begin{centering}
\includegraphics[trim = 0 0 0 0, clip, width=0.99\textwidth]{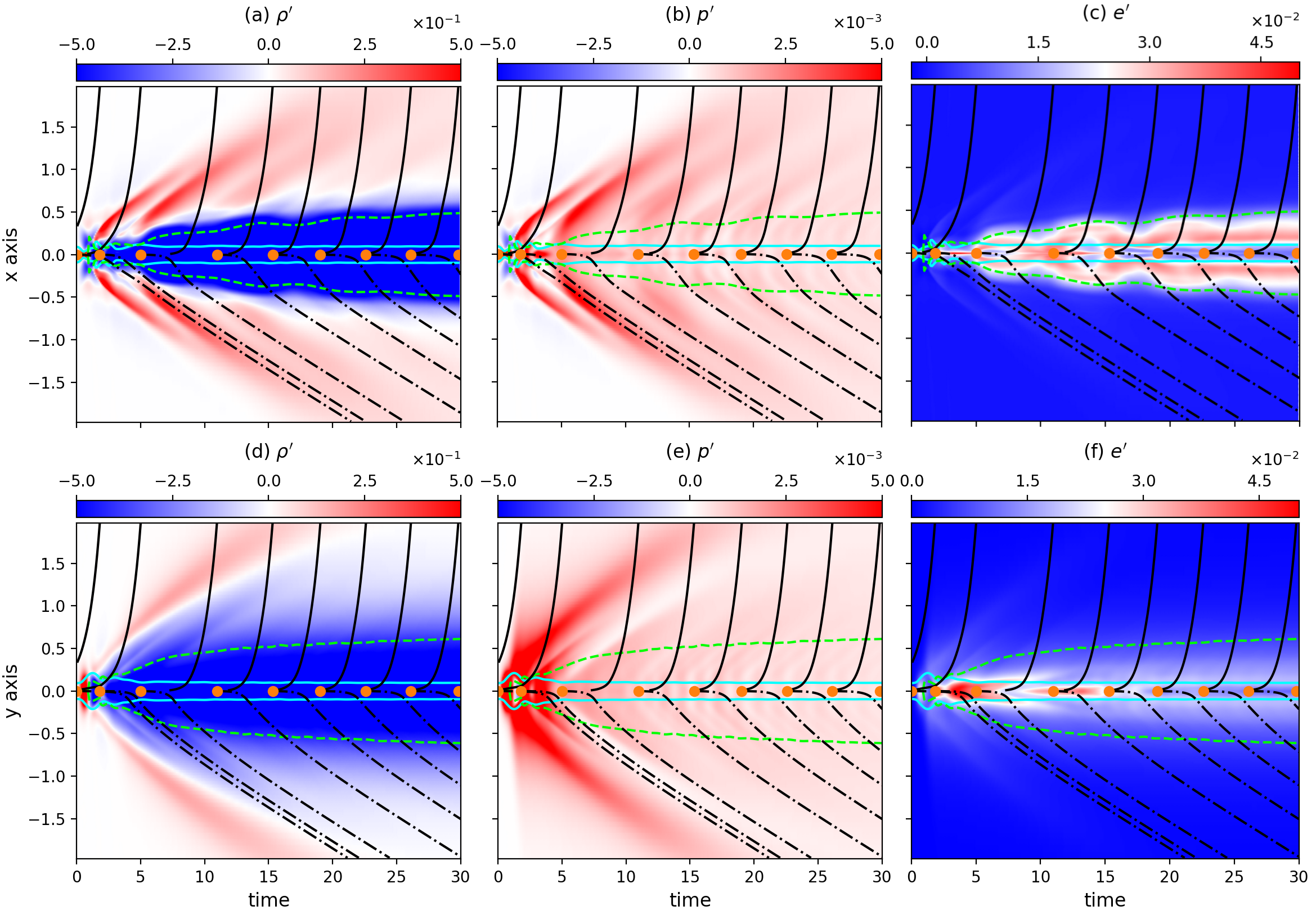}
\end{centering}
\caption{Time-distance diagrams showing $\rho^\prime$, $p^\prime$, and $e^\prime$ perturbations along the $x$-axis and $y$-axis. The orange dots mark the roots of $j_y(0,0,0,t)$, while cyan lines indicate the equipartition layer and green lines the cavity envelope. The black dashed-dotted and black continuous lines represent the slow and fast magnetoacoustic wave trajectories respectively. }
	\label{fig:st-thermo-fan}
\end{figure*}

\subsubsection{Wave propagation along the spine}

\begin{figure*}[t]
\includegraphics[trim = 0 0 0 0, clip, width=0.99\textwidth]{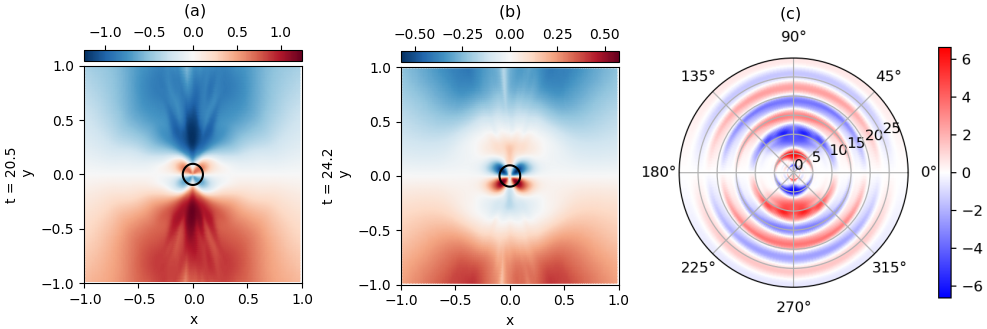}
	\caption{(a) and (b) are contour plots in the $z=0$ plane displaying the time evolution of the $\xi_A$. The black circles represent the equipartition layer. The simulation time instants shown here were selected from the orange dots in Figure \ref{fig:jy}. (c) Polar time-distance diagram for $\xi_A$ in the fan plane where the radial direction represents the simulation time. $\xi_A$ is monitored at a circular line with coordinates $(r=0.1,\theta,z=0,t)$.}
	\label{fig:fan-MHD-alfven}
\end{figure*}

After characterizing OR ($j_y$ at the null point) and wave propagation patterns along the spine, we move the focus to MHD wave types. Two complementary approaches are used to identify the waves: time-distance diagrams tracking perturbation evolution along characteristic field lines for the wave proxies, and the wave tracers discussed in \S \ref{sec:tracing}. 

Regarding the analysis of the MHD wave proxies, Figure \ref{Figure 8 new} presents present time-distance diagrams for compressible parallel and transverse wave proxies, $\xi_\|$ and $\xi_\bot$, with their SPOD reconstructions, $\widetilde{\xi}_\|$ and $\widetilde{\xi}_\bot$. $\xi_A$ (incompressible, Alfvén-like waves) were omitted from Figure \ref{Figure 8 new}, since no signal was found propagating along the spine. \new{Here we use pairs of modes for reconstructing propagating waves. According to \citep{taira_modal_2017} POD and SPOD modes that are real do not contain information of the wave phase (since there is no imaginary part). Instead, they require the superposition of a pair of modes with a phase shift for describing any propagating wave.}

The first proxy $\xi_\|$ in Figure \ref{Figure 8 new}a shows compressible perturbations parallel to the magnetic field. There is an initial transient from $t=0$ to 10, and approximately after the middle of the second reconnection cycle, the propagation (at the third orange dot) starts to develop fully.  Additionally, there is no information traveling at the fast speed (black continuum lines) in $\xi_\|$.

Figures \ref{Figure 8 new}b and \ref{Figure 8 new}c present SPOD reconstructions of $\xi_\|$ using modes 1-2 and modes 3-4, respectively. The modes 1-2 reconstruction (Figure \ref{Figure 8 new}b) reveals a wave traveling with the slow speed (dashed-dot lines) and starts at the $j_y(0,0,0,t)$ roots (shown as orange dots, i.e. the roots  in Figure \ref{fig:jy}a) with period $P$. 

The modes 3-4 reconstruction (Figure \ref{Figure 8 new}c) identifies previously unobserved short-period waves, with periodicity $P/2$, with a significant interference pattern effect (within the range $ z \approx \pm 0.5$). 
It is unclear if these are standing or propagating waves, due to the strong interference pattern. Outside of $z \approx \pm 0.5$, the periodicity becomes $P$ and the waves propagate at the fast magnetoacoustic speed. These features were absent in the original diagram (Figure \ref{Figure 8 new}a) which demonstrates the utility of SPOD.

Figure \ref{Figure 8 new}d shows compressible transverse perturbations by monitoring $\xi_\bot$, and Figures \ref{Figure 8 new}e and \ref{Figure 8 new}f present SPOD reconstructions of $\xi_\bot$ using modes 1-2 and modes 3-4, respectively. Figure \ref{Figure 8 new}d shows that, during the initial transient, from $t=0$ to 10, there are compressible transverse waves with period $P/2$ traveling with the fast speed (black continuum lines) along the spine, which are not seen after $t=10$.

The reconstruction using modes 1-2 (shown in Figure \ref{Figure 8 new}e) reveals propagating waves, traveling at the slow speed, with period $P$, i.e. synchronized with the $j_y(0,0,0,t)$ roots (Figure \ref{fig:jy}a).

In contrast, the modes 3-4 reconstruction (Figure \ref{Figure 8 new}f) uncovers shorter period, $P/2$, standing waves, that decay rapidly at $z \approx \pm 0.5$. Waves outside this are rapidly damped, i.e. outside $z \approx \pm 0.5$.

In order to explain the bounding of the standing waves in Figure \ref{Figure 8 new}c and \ref{Figure 8 new}f), Figure \ref{fig:st-thermo-spine} shows the time-distance diagrams along the spine for $\rho^\prime$, $p^\prime$ and $e^\prime$.

\new{Figure \ref{fig:st-thermo-spine}a shows the propagation of waves at the slow speed. From $t=5$ onward a density depletion forms and move at the slow speed along the spine to approximately $\pm1$. Later, from $t=10$ onward, a series of depletions form at the null and move to $\pm0.5$.
A similar behavior is seen in Figure \ref{fig:st-thermo-spine}b for $p^\prime$. Notice that there is no signature of this in the perturbed energy in Figure \ref{fig:st-thermo-spine}c. These density and pressure depletions provide the inhomogeneity in the plasma medium that cause the standing waves seen in Figure \ref{Figure 8 new}c and \ref{Figure 8 new}f, as well as the waves traveling at the slow speed in Figure \ref{Figure 8 new}a and \ref{Figure 8 new}b.}

%
%

\subsection{Waves in the fan plane} \label{sec:waves-fan}
\subsubsection{The cavity}

Our investigation now turns to the fan plane, which is approximately located at $z=0$ and oscillates around this position (Figure \ref{fig:jy}c). As opposed to the spine, the field line tracing in the fan plane is more complex because they diverge at the null point and are not anchored to any particular position during the system's evolution.

Figure \ref{fig:slices-fan} displays contour plots of $\rho^\prime$ and $p^\prime$ that represent evolution along the $z=0$ plane, which is approximately the fan plane. At $t=0.2$, the initial transient generates pressure and density disturbances that travels towards the null point. Following the arrival of the initial pulse at the null point, as can be seen at $t=6.8$, a circular wave is emitted within the fan plane ($xy$-plane, $z=0$). The first wavefront, associated with the initial perturbation, is strongest along the $y$-direction and is located near $x=0$, $y=\pm1$, 
and a second wavefront at $y = 0$, $x = \pm 0.5$, where both waves disturbs the density and pressure.
Later in time, at $t=17.1$, wavefronts are observed near $x=\pm0.5$, $y=0$ that exhibit stronger amplitudes in $p^\prime$ along the $x$-direction and also disturb the density.

After $t=17.1$, the wave pattern in the fan plane persists but is now predominant along the $x$-direction. Subsequent snapshots at $t=20.5$ and $t=24.1$ reveal a similar pattern in $p^\prime$, though with a reduced amplitude.
The influence of pressure perturbations on the density after $t=6.8$ is not visible in the $\rho^\prime$ contour. After $t=6.8$ the evolution produces a significant density depletion within the fan plane. This depletion is localized to a region approximately within $\sqrt{x^2+y^2}<0.5$ and is non-axisymmetric, resulting in a density drop of 70–80\%. This indicates the generation of a cavity inside the fan plane ($z = 0$) at $\sqrt{x^2+y^2}<0.5$.

The evolution of the thermodynamic variables are analyzed in more detail in the fan plane in Figure \ref{fig:st-thermo-fan}. The dashed-dotted lines in Figure \ref{fig:st-thermo-fan} represent tracers based on the slow magnetoacoustic speed, while the solid lines correspond to the fast magnetoacoustic speed. A green dashed line denotes the boundary of the evolving cavity. This cavity profile was determined by applying a Gaussian fit to the density distribution near the null point at each time step of the time-distance diagram, the cavity boundary is defined as the standard deviation of the Gaussian function.

Along the $x$-axis (Figures \ref{fig:st-thermo-fan}a-\ref{fig:st-thermo-fan}c), there is a sharp drop in density perturbation within the cavity region (Figure \ref{fig:st-thermo-fan}a) and wave generation occurs outside this cavity shell. In contrast, the pressure perturbation diagram (Figure \ref{fig:st-thermo-fan}b) shows perturbations that originate at the null point. \new{Figure \ref{fig:st-thermo-fan}c highlights that perturbed internal energy is predominantly confined within the cavity, thus heating the plasma around the null point}. Figures \ref{fig:st-thermo-fan}d-\ref{fig:st-thermo-fan}f exhibit similar behavior along the $y$-axis, albeit with some differences: the density profile transitions more smoothly across the cavity boundary (Figure \ref{fig:st-thermo-fan}d) compared to the $x$-axis (Figure \ref{fig:st-thermo-fan}a), which is consistent with the earlier analysis in Figure \ref{fig:slices-fan}. The pressure perturbations (Figure \ref{fig:st-thermo-fan}e) show a pattern similar to that observed along the $x$-axis. However, the internal energy perturbations (Figure \ref{fig:st-thermo-fan}f) reveal a notable contrast. The heating along the $y$-axis is less intense since the \new{reconnection jets are aligned with} to the $y=0$ plane and remain stronger within the equipartition layer. It also display a peak of heat at $t=7$, after the first reconnection cycle. In contrast, heating along the $x$-axis is more widely distributed throughout the cavity.

%
%

\subsubsection{Incompressible waves ($\xi_A$)}

\begin{figure*}[t]
\centering
\includegraphics[trim = 0 0 0 0, clip, width=0.99\textwidth]{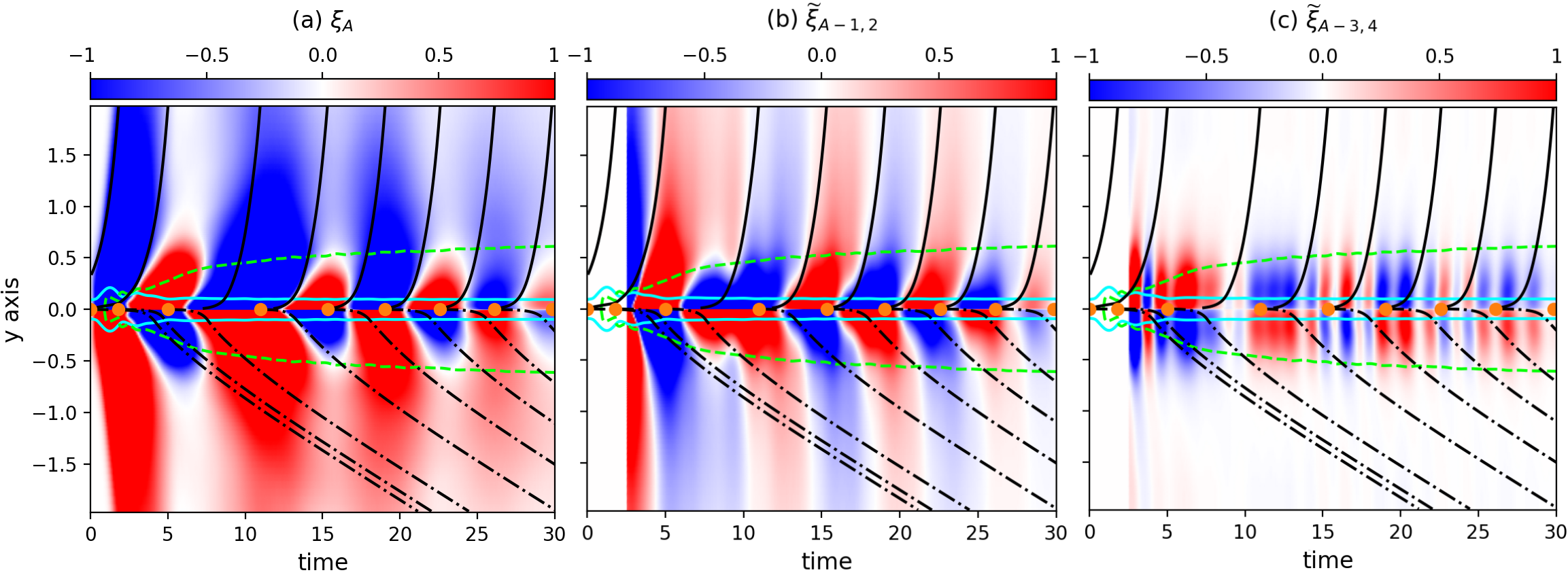}
\caption{Time-distance diagrams showing for (a) $\xi_A$ and (b) and (c) its SPOD reconstructions, $\widetilde{\xi}_A$, along the $y$-axis. The orange dots mark the roots of $j_y(0,0,0,t)$ in Figure \ref{fig:jy}a, while cyan lines indicate the equipartition layer and green lines the cavity envelope. The black dashed-dotted and black continuous lines represent the slow and fast magnetoacoustic wave trajectories respectively. }
	\label{fig:st-fan-alfven}
\end{figure*}

Figure \ref{fig:fan-MHD-alfven}a-b illustrates the time evolution of the MHD Alfvén wave proxy $\xi_A$ near the fan plane, at $z = 0$, and
are shown over two reconnection cycles, at $t =$ 20.5 and 24.2, corresponding to the last two orange dots in the $j_y(0,0,0,t)$ time series (Figure \ref{fig:jy}a).

\new{The evolution of $\xi_A$ indicates the presence of an incompressible field-aligned vorticity propagating in the fan plane along the $y$-axis.} This wave is perpendicular to the spine motion, which occurs in the $xz$-plane at $y = 0$ (see Figure \ref{fig:jy}b). The bending of the fan plane at $y = 0$ (as shown in Figure \ref{fig:jy}c) generates vorticity that aligns with the magnetic field, resulting in $\xi_A$. The oscillation period of $\xi_A$ appears to be synchronized with $j_y(0,0,0,t)$, supporting the conclusion that $\xi_A$ originates from the bending of the fan plane. Notably, the wave pattern appears to be influenced by the cavity at $t=24.2$, where the pattern is contained at boundary defined by $\sqrt{x^2 + y^2} \approx 0.5$.

Figure \ref{fig:fan-MHD-alfven}c presents a polar time-distance diagram for $\xi_A$ in the fan plane, specifically for a circular path at a constant radius, $\xi_A(r=0.1,\theta, z=0, t)$. The contour plot indicates that there is no propagation of incompressible waves along the $x$ axis ($\theta=0^o$ and 180$^o$). Instead, wave propagation is predominantly along the $y$ axis ($\theta=90^o$ and 270$^o$), which aligns with the direction of the current density vector at the null ($\mathbf{j}= (0,j_y,0)$ see Figure \ref{fig:jy}a). The polar time-distance diagram shows a dipole pattern and agrees with the behavior seen in Figure \ref{fig:fan-MHD-alfven}a-b. Note that $j_y(0,0,0,t)$ in Figure \ref{fig:jy}a changes sign as OR evolves in time and this pattern is repeated by $\xi_A$ inside the equipartition layer, as can be seen in Figures \ref{fig:fan-MHD-alfven}a-b.

Figures \ref{fig:st-fan-alfven}a-\ref{fig:st-fan-alfven}c present time-distance diagrams along the $y$-axis for $\xi_A$, as well as its SPOD reconstructions of $\widetilde{\xi}_A$ using modes 1-2 and 3-4. These SPOD modes are used to study wave signals and disentangle the complex dynamics into fundamental components, aiding in identifying the origin of each wave mode in the system. In Figure \ref{fig:st-fan-alfven}, the continuous lines represent the trajectories of the fast speed, while the dashed-dot lines indicate the trajectories of the slow speed.

Figure \ref{fig:st-fan-alfven}a shows that a wave exists in the system, propagating mainly in the $y$-direction at the fast speed. The wave signal originates at the null point and has the same period as $j_y(0,0,0,t)$, i.e. a period of $P$. Magnetic reconnection creates movement of the spine in the $y=0$, $xz$-plane, which results in the bending of the fan plane (Figure \ref{fig:jy}c). This bending generates an incompressible wave that propagates transverse to the spine movement in the fan plane along the $y$-direction. Moreover, the incompressible waves are observed to be more intense within the cavity, and some pulses appear to be contained inside of it.

Figures \ref{fig:st-fan-alfven}b and \ref{fig:st-fan-alfven}c provide SPOD reconstructions for the incompressible wave proxy. A comparison between the original data (Figure \ref{fig:st-fan-alfven}a) and the modes 1-2 reconstruction (Figure \ref{fig:st-fan-alfven}b) reveals a previously obscured wave propagation beyond the cavity (indicated by the green dashed lines) with a period of $P$. These waves originate at the null point, exhibit maximum amplitude within the cavity, and propagate outward with diminishing intensity. The modes 3-4 reconstruction (Figure \ref{fig:st-fan-alfven}c) shows standing waves inside the cavity with periods $P/2$. These standing waves also propagate outward across the cavity boundary but then experience rapid amplitude decay.

\begin{figure*}[t]
\centering
\includegraphics[trim = 0 0 0 0, clip, width=0.99\textwidth]{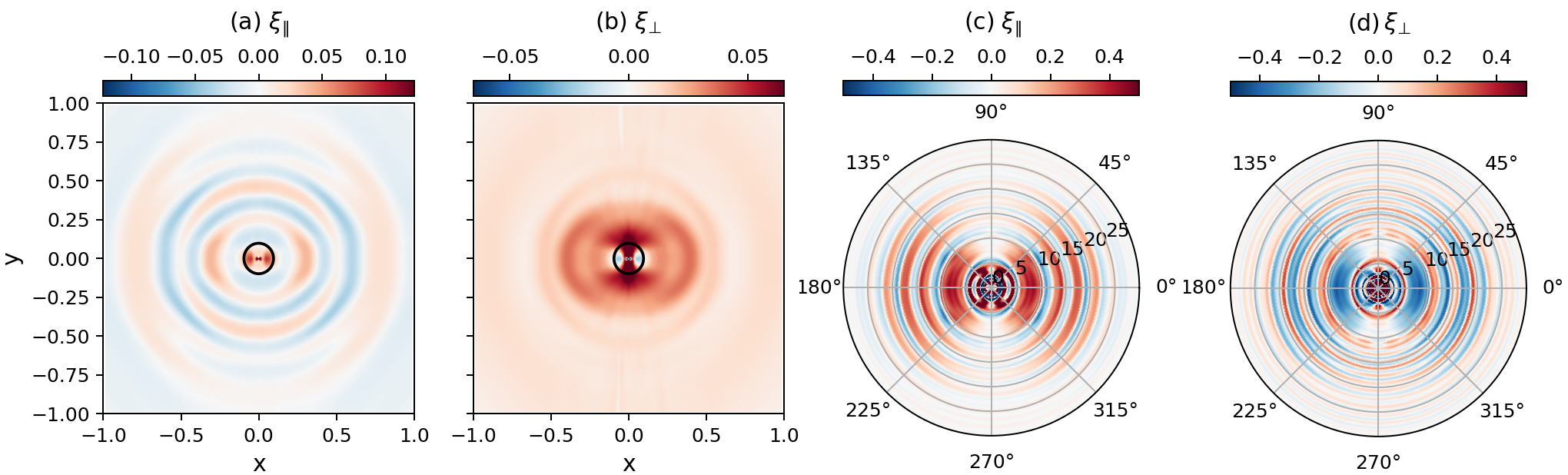}
\caption{Panels (a) and (b) show the compressible wave proxies in the fan plane, $z=0$, at $t=24.2$. The black circle represent the equipartition layer. Panels (c) and (d) show a polar time-distance plot for the compressible wave proxies obtained at $(r=0.1,\theta,z=0,t)$, where the radial direction represents the simulation time. }
\label{fig:proxies-fan-compressible}
\end{figure*}

%
%

\subsubsection{Compressible waves in the fan plane} 
$\xi_\|$ and $\xi_\bot$ are employed to investigate compressible wave motion in the fan plane. Figure \ref{fig:proxies-fan-compressible} illustrates the wave patterns for $\xi_\|$ and $\xi_\bot$ near the fan plane, specifically at $z = 0$ and $t = 24.2$, after two reconnection cycles, and a polar time distance diagram for both variables.

Figure \ref{fig:proxies-fan-compressible}a contains contour plots of $\xi_\|(x,y,0,t=24.2)$. This contour plot shows that circular compressible waves exist outside the equipartition layer (i.e., where $\beta < 1$). Outside this layer, a wave pattern propagates radially across the fan plane at $z = 0$. Three distinct patterns are observable in this case. While some dynamics occur within the equipartition layer (where $\sqrt{x^2 + y^2} \lesssim 0.1$), the pattern is not clearly defined in the contour plot. Inside the equipartition layer it is seen that $\xi_\|$ has the same sign of $j_y(0,0,0,t=24.2)$ (Figure \ref{fig:jy}a). Beyond this layer however, a well-defined wave pattern emerges, with wave propagation observed up to $\sqrt{x^2 + y^2} \approx 0.5$, which is approximately the cavity boundary, beyond which their wavelength increases. This behavior is mainly created by the perturbations in the fan plane shown in Figure \ref{fig:POD-modes3D}b.

Figure \ref{fig:proxies-fan-compressible}b presents a contour plot of $\xi_\bot(x,y,z=0,t=24.2)$. Here, $\xi_\bot(x,y,0,24.2)$ exhibits a circular wave signal between the equipartition layer and the cavity boundary (where $\sqrt{x^2 + y^2} \approx 0.5$), which may be attributed to fast magnetoacoustic waves. The amplitudes of these waves are smaller than those of $\xi_\|$ and 10 times smaller than $\xi_A$. Also from the contour panel, it is unclear whether waves from $\xi_\bot$ propagate beyond the cavity or if their amplitude diminishes, rendering them undetectable. Inside the equipartition layer, it is observed that $\xi_\bot$ shares the opposite sign as $j_y(0,0,0,t=24.2)$ (Figure \ref{fig:jy}a) while $\xi_\|$ has the same sign in the same region.

\begin{figure*}[t]
\centering
\includegraphics[trim = 0 0 0 0, clip, width=0.99\textwidth]{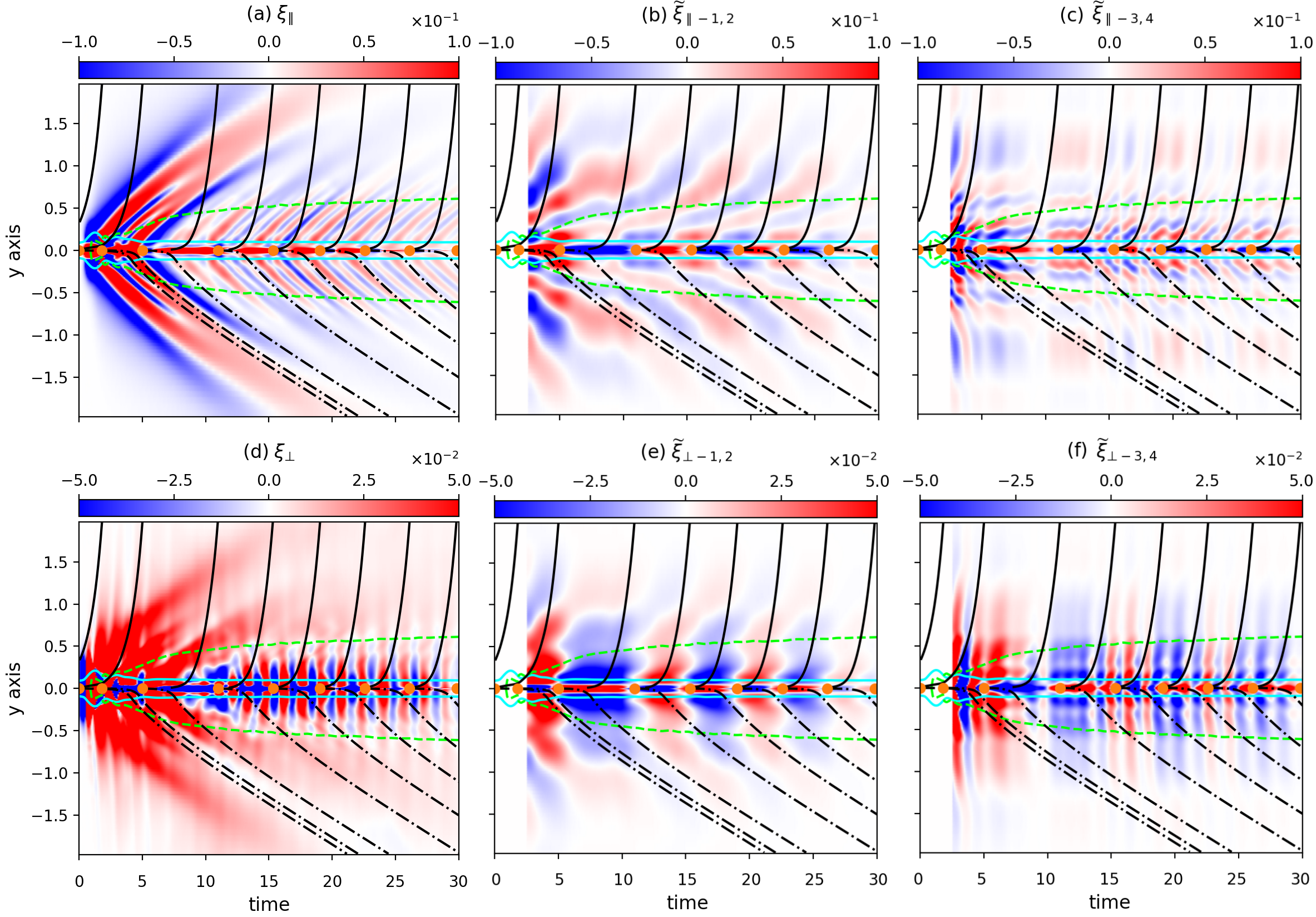}
\caption{Time-distance diagrams for $\xi_\|$ and $\xi_\bot$ and their SPOD reconstruction along the $y$-axis. The orange dots mark the roots of $j_y(0,0,0,t)$, while cyan lines indicate the equipartition layer and green lines the cavity envelope. The black dashed-dotted and black continuous lines represent the slow and fast magnetoacoustic wave trajectories respectively. }
	\label{fig:st-proxies-y}
\end{figure*}

Figures \ref{fig:proxies-fan-compressible}c and \ref{fig:proxies-fan-compressible}d present polar time-distance diagrams for  $\xi_\|$ and $\xi_\bot$ along a line at $\sqrt{x^2 + y^2} = 0.1$, outside the equipartition layer and inside the cavity, and $z = 0$. Figure \ref{fig:proxies-fan-compressible}c shows that $\xi_\|$ exhibits a circular waves that are non-uniform in the polar direction; it oscillates between -0.5 and 0.5, with a reduction in the amplitude along the $y-$axis 
(at $\theta = 90^o$ and $270^o$), where it oscillates between -0.2 and 0.2. 
The oscillation amplitudes are higher along the $x-$axis (at $\theta = 0^o$ and $180^o$).
In figure \ref{fig:proxies-fan-compressible}d, $\xi_\bot$ displays a circular wave pattern where the wave amplitude is higher along the $x-$axis (at $\theta = 0^o$ and $180^o$). The propagation along the $y-$axis (at $\theta = 90^o$ and $270^o$) are observed but are less intense, showing only positive values, while along the $x-$axis it oscillates between -0.5 to 0.5.

\begin{figure*}[t]
\includegraphics[trim = 0 0 0 0, clip, width=0.99\textwidth]{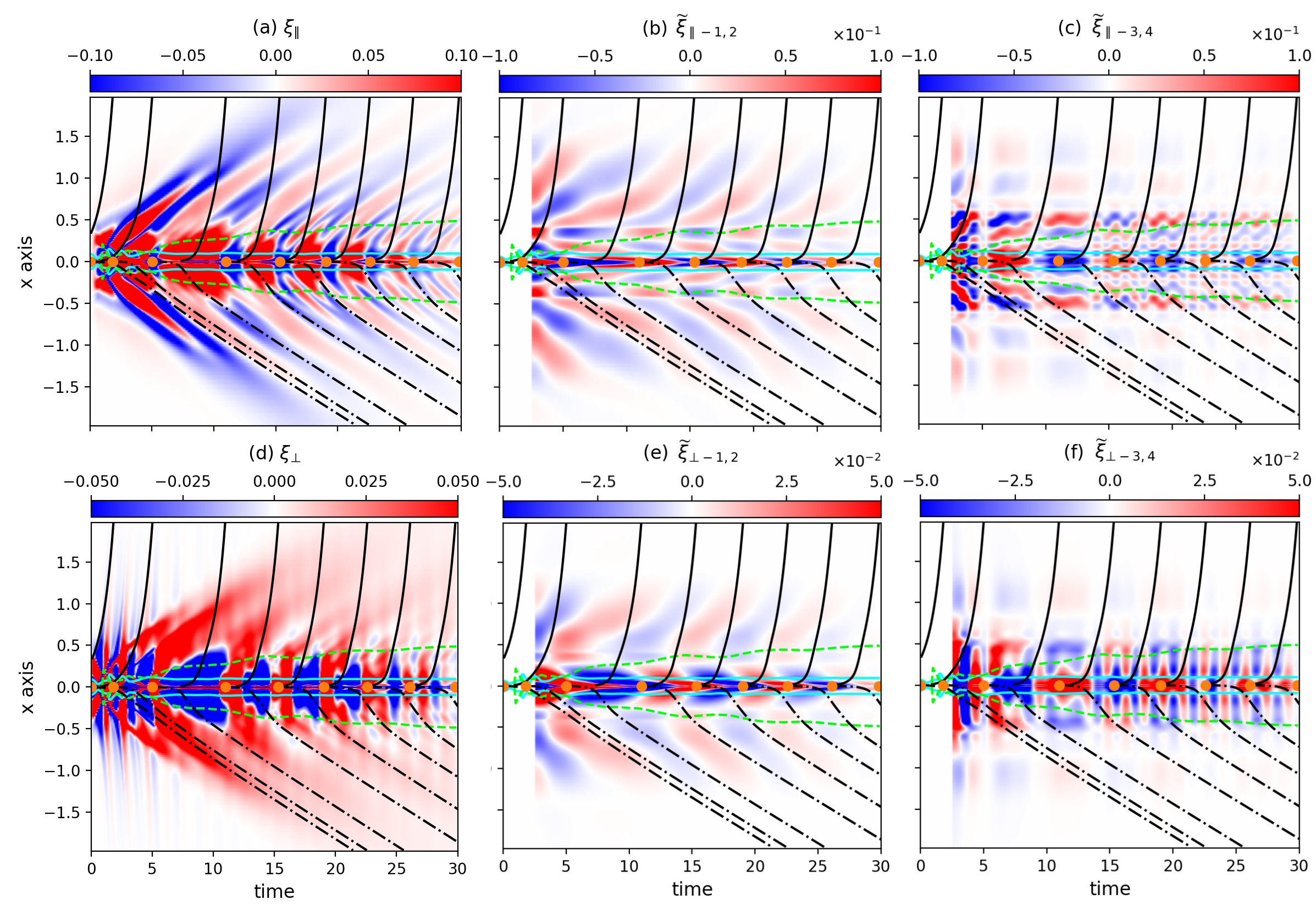}
\caption{Time-distance diagrams for $\xi_\|$ and $\xi_\bot$ and their SPOD reconstruction along the  $x$-axis. The orange dots mark the roots of $j_y(0,0,0,t)$, while cyan lines indicate the equipartition layer and green lines the cavity envelope. The black dashed-dotted and black continuous lines represent the slow and fast magnetoacoustic wave trajectories respectively. }
	\label{new Figure 15}
\end{figure*}

\subsubsection{Compressible waves along the $y$-axis}\label{Sec:3.3.4}


The next analysis moves to the evolution along the $y$-axis in the fan plane. Figures \ref{fig:st-proxies-y}a to \ref{fig:st-proxies-y}f present time-distance diagrams for compressible parallel and transverse wave proxies, $\xi_\|$ and $\xi_\bot$, with their SPOD reconstructions, $\widetilde{\xi}_\|$ and $\widetilde{\xi}_\bot$. 

Figure \ref{fig:st-proxies-y}a illustrates the compressible parallel perturbations, $\xi_\|$, along the $y$-axis in the fan plane. Between $t=0$ and 10, there is an initial transient and the formation of the low-density cavity (dashed green line). After the initial transient, short-period oscillations within the cavity propagate at a slow speed, at a period of approximately $P/2$, and terminate at the cavity boundary. Outside the cavity, there are compressible waves that are synchronized with  $j_y(0,0,0,t)$ roots in Figure \ref{fig:jy}a with a period of $P$ and traveling at the slow speed (following the dashed-dot paths). This confirms the results in Figure \ref{fig:proxies-fan-compressible}a. 

Figures \ref{fig:st-proxies-y}b and \ref{fig:st-proxies-y}c present SPOD reconstructions using modes 1-2 and modes 3-4, respectively. The modes 1-2 reconstruction (Figure \ref{fig:st-proxies-y}b) confirms the synchronization with the $j_y(0,0,0,t)$ roots in Figure \ref{fig:jy}a and verifies the propagation of slow magnetoacoustic waves beyond the cavity.

Figure \ref{fig:st-proxies-y}c shows the modes 3-4 reconstruction. Withing the cavity, short-period waves with periodicity $P/2$ are observed, propagating at the local slow speed. Outside the cavity, these waves propagate at the fast magnetoacoustic speed with period $P$.

Figure \ref{fig:st-proxies-y}d, shows the transverse compressible perturbations in $\xi_\bot$. The transverse perturbations exhibit behavior similar to that seen along the spine (as shown in Figure \ref{Figure 8 new}d). During the initial transient, $t = 0$ to $t = 10$,
oscillations are displaying shorter periods than the OR period, specifically a period of $P/2$ between $y\pm0.5$ to 2 (following fast magnetoacoustic speed black lines).

Figures \ref{fig:POD-modes3D}c and \ref{fig:proxies-fan-compressible}b emphasize that the perturbation is contained mainly inside the cavity.

Figures \ref{fig:st-proxies-y}e and \ref{fig:st-proxies-y}f compare SPOD reconstructions using two different combinations of modes: modes 1-2 and modes 3-4. The reconstruction using modes 1-2 (shown in Figure \ref{fig:st-proxies-y}e) reveals propagating waves synchronized with the $j_y(0,0,0,t)$ roots (Figure \ref{fig:jy}a) within and outside the cavity.

In contrast, the modes 3-4 reconstruction (Figure \ref{fig:st-proxies-y}f) uncovers $P/2$ cavity-trapped, standing oscillations. Waves outside the cavity (i.e. outside $ y = \pm 0.5$)  propagate at the fast magnetoacoustic speed (following the black continuous lines) but they decay rapidly.

\subsubsection{Compressible waves along the $x$-axis} 

\new{Finally, the compressible wave proxies along the $x-$axis are investigated. The results can be seen in Figure \ref{new Figure 15}, which shows the time-distance diagrams for  $\xi_\|$ and $\xi_\bot$ and their SPOD reconstruction along the  $x$-axis. The results are very similar to those in $\S\ref{Sec:3.3.4}$, for example Figure \ref{new Figure 15}b  reports a  wave traveling with the slow speed (dashed-dot lines)  with period $P$, and thus the behavior is reminiscent of that observed for the same variable along the spine (Figure \ref{Figure 8 new}b) and $y-$axis (Figure \ref{fig:st-proxies-y}b). Thus, the full breakdown of the results in  Figure \ref{new Figure 15} is not repeated here.
}

\section{Conclusions} \label{sec:conclusions}

The investigation of wave generation at a 3D magnetic null point reveals several key insights into the coupling between oscillatory reconnection (OR) and magnetohydrodynamic (MHD) wave dynamics. The analysis demonstrates that OR serves as a robust mechanism for generating self-sustained oscillations, even when initiated by a single aperiodic perturbation. The periodic reversals in current density at the null point, synchronized with fan plane deformations, drive a rich spectrum of MHD waves across the spine and fan structures.


The OR signature is quantified by $j_y(0,0,0,t)$.
Its roots (Figure \ref{fig:jy}a) determines a characteristic period, $P$, in the system. Wave motions of period $P$ and $P/2$ are observed in the system (see below). $j_y(0,0,0,t)$ exhibits a Gaussian decay, as discussed in \cite{schiavo2025OR}. It was tracked via the motion of the spine and fan plane in the $xz-$plane at $y=0$. It was demonstrated that both the spine and fan are deformed, with the perturbation amplitude decaying significantly after the first reconnection cycle. The maximum amplitude decrease occurs in the first OR cycle, approximately 6 times, representing a six-fold reduction (Figure \ref{fig:jy}c). Additionally, the OR perturbation continually deforms the fan plane.


This investigation utilized the  approach from \cite{Raboonik2022} to isolate an incompressible parallel component, $\xi_A$, a compressible parallel component, $\xi_{\|}$, and a compressible transverse component, $\xi_{\perp}$. These identifiers are not exact representations of the three fundamental MHD wave modes, but they encapsulate the essential features distinguishing wave behavior, and hence are useful in wave mode identification. This investigation also utilized Spectral Proper Orthogonal Decomposition (SPOD) to analyze the wave motions. The first four SPOD modes captured approximately 90\% of the system's fluctuation energy (Figure \ref{fig:pod-energy}). Mode 1 accounted for about 73\% and was predominantly influenced by the OR period. It was found that modes 1 and 2, which together contain approximately 80\% of the fluctuation energy, were sufficient to describe the incompressible and compressible magnetoacoustic waves that share the same OR period, $P$. In contrast, modes 3 and 4, which account for around 10\% of the fluctuation energy, were responsible for describing the waves with half the OR period ($P/2$), including compressible parallel and transverse, and incompressible waves (Figures \ref{fig:st-fan-alfven} and \ref{fig:st-proxies-y}).


The plasma heating generated by the reconnection jets creates a cavity in the fan plane, characterized by a region of reduced density near the null point (see Figures \ref{fig:slices-fan}, \ref{fig:st-thermo-fan}a, \ref{fig:st-thermo-fan}d). This cavity plays a crucial role in the dynamics of the system: It traps heat, as evidenced by internal energy perturbations (see Figures \ref{fig:st-thermo-fan}c, \ref{fig:st-thermo-fan}f), and modulates wave propagation. Separately, density depletions in the equipartition layer along the spine were reported (Figure  \ref{fig:st-thermo-spine}). This created a region of trapped waves at $z \approx \pm 0.5$ along the spine, allowing the manifestation of standing waves of period $P/2$.


Wave motions along the spine ($z-$axis) and fan plane ($x-$ and $y-$axes) were investigated. In all three directions, common wave behavior was reported:
\begin{enumerate}[(i)]
    \item {There exists an initial transient between $t=0$ and $t=10$. Compressible transverse waves ($\xi_\bot$) propagating with the fast speed and with period $P/2$ were observed along the spine as well as the fan plane along the $x-$ and $y-$axes  (Figures \ref{Figure 8 new}d,  \ref{fig:st-proxies-y}d and \ref{new Figure 15}d). }
    \item{For each of the 3 directions, there always exists a period $P$ wave, propagating at the local slow speed (both inside and outside the spine equipartition and fan cavity). This manifests in both $\xi_\parallel$ and $\xi_\bot$, along the spine (Figures \ref{Figure 8 new}b and \ref{Figure 8 new}e), along the $y-$axis (Figures \ref{fig:st-proxies-y}b and \ref{fig:st-proxies-y}e) and along the $x-$axis (Figures \ref{new Figure 15}b and \ref{new Figure 15}e). Furthermore, Figures \ref{fig:proxies-fan-compressible}a and \ref{fig:proxies-fan-compressible}b reveal that a propagating, period $P$, slow wave occurs in all fan-plane directions. Figure \ref{fig:proxies-fan-compressible}c showed that period $P$ was axisymmetric in the fan plane, with the amplitude of such slow waves stronger in the $x-$direction (maximum amplitude $\approx 0.5$) compared to the $y-$direction (maximum amplitude $\approx 0.2$).}
    \item{For each of the 3 directions, there always exists a second wave of period $P/2$, characterized by being trapped by density depletions along the spine (Figure \ref{fig:st-thermo-spine}a) and inside the fan cavity (Figures \ref{fig:st-thermo-fan}a and \ref{fig:st-thermo-fan}d). This wave manifests as a standing wave in $\xi_\parallel$ and $\xi_\bot$, along the spine (Figures \ref{Figure 8 new}c and \ref{Figure 8 new}f), along the $y-$axis (Figures \ref{fig:st-proxies-y}c and \ref{fig:st-proxies-y}f) and along the $x-$axis (Figures \ref{new Figure 15}c and \ref{new Figure 15}f). Outside the density depletions along the spine and the fan cavity, different behavior is reported: a propagating fast wave with period $P$ is observed in $\xi_\parallel$ (Figures \ref{Figure 8 new}c, \ref{fig:st-proxies-y}c and \ref{new Figure 15}c), whereas wave behavior in $\xi_\bot$ decays rapidly in amplitude (Figures \ref{Figure 8 new}f, \ref{fig:st-proxies-y}f and \ref{new Figure 15}f).} 
\end{enumerate}


For the initial transient (propagating wave of period $P/2$), these are compressible transverse waves ($\xi_\bot$) propagating with the fast speed, and so we interpret these as fast magnetoacoustic waves. For the propagating wave of period $P$ that manifests along both the spine and the whole fan plane, these are compressible  waves ($\xi_\parallel$) propagating parallel to the magnetic field lines, traveling at the slow speed, and so we interpret these as slow magnetoacoustic waves.

These results also demonstrate the utility of the SPOD reconstruction: SPOD modes revealed previously-obscured wave propagation in all directions and for all three MHD waves. Note that point (ii) is derived from modes 1 and 2, which together contain approximately 80\% of the fluctuation energy, and point (iii) is related to information derived from modes 3 and 4, which account for around 10\% of the fluctuation energy.



Along the spine and the $x$-direction of the fan plane (see Figures \ref{Figure 8 new} and \ref{new Figure 15}), only waves that disturb the density, i.e. magnetoacoustic, were found; the incompressible wave proxy $\xi_A$ was near zero. Incompressible waves were exclusively detected along the $y$-axis in the fan plane, propagating transverse to the plane containing the spine motion
(Figures \ref{fig:fan-MHD-alfven} and \ref{fig:st-fan-alfven}). They have period $P$, synchronized with the roots of $j_y(0,0,0,t)$, and propagate both inside and outside of the cavity at the local fast speed (Figure \ref{fig:st-fan-alfven}b). The local fast speed is approximately the same as the local Alfv\'en speed in this region, and  these period $P$ incompressible waves propagate at the local Alfv\'en speed. They also propagate vorticity (due to the formalism of $\xi_A$, Equation (\ref{eq:alfven})) and thus we interpret these as a propagating Alfv\'en wave.

SPOD analysis also revealed low-amplitude, $\xi_A$ standing waves with a period of  $P/2$ inside the cavity. These rapidly decay outside the cavity (Figure \ref{fig:st-fan-alfven}c).

This work demonstrates that a 3D null point can function as a self-oscillating wave source, with its wave properties intrinsically linked to $j_y$ oscillations at the null point. An aperiodic disturbance leads to the generation of propagating, period $P$, slow magnetoacoustic waves, along both the spine and the fan-plane. For a 2D X-type null point undergoing OR, \citet{Karampelas2023} has shown a direct link between the OR period $P$ and the magnetic field strength, density and temperature (i.e. a dependence upon the local Alfv\'en speed and local sound speed). No such parameter study has been conducted for a 3D null. However, should the 2D results of \citet{Karampelas2023} translate to 3D, then the  propagating, period $P$, slow magnetoacoustic waves reported in this paper open up a fascinating  opportunity for indirectly deriving information about the local magnetic field strength, density and temperature, i.e. new avenues for coronal seismology.


\section*{Acknowledgments}
All authors acknowledge the UK Research and Innovation (UKRI) Science and Technology Facilities Council (STFC) for support from grant ST/X001008/1 and for IDL support. The research was sponsored by the DynaSun project and has thus received funding under the Horizon Europe programme of the European Union under grant agreement (no. 101131534). Views and opinions expressed are however those of the author(s) only and do not necessarily reflect those of the European Union and therefore the European Union cannot be held responsible for them. This work was also supported by the Engineering and Physical Sciences Research Council (EP/Y037464/1) under the Horizon Europe Guarantee. This work used the Oswald High Performance Computing facility operated by Northumbria University (UK), and the DiRAC Data Intensive service (CSD3) at the University of Cambridge, managed by the University of Cambridge University Information Services on behalf of the STFC DiRAC HPC Facility (www.dirac.ac.uk). The DiRAC component of CSD3 at Cambridge was funded by BEIS, UKRI and STFC capital funding and STFC operations grants. DiRAC is part of the UKRI Digital Research Infrastructure. Numerical simulations were conducted with LARE3D which is available at \href{https://github.com/Warwick-Plasma/Lare3d}{https://github.com/Warwick-Plasma/Lare3d}. \new{The data behind figures and the inputs from the simulation are available in Figshare at DOI:\href{https://doi.org/10.25398/rd.northumbria.30896033}{https://doi.org/10.25398/rd.northumbria.30896033}. Additional data data support the findings of this study are available from the corresponding author upon reasonable request.}

\software{LARE3D \citep{Arber2001},
          NumPy \citep{numpy},  
          SciPy \citep{SciPy}, 
          Matplotlib \citep{matplotlib}
          WaLSA tools \citep{jafarzadeh_wave_2025}.
          }


\bibliography{references}{}
\bibliographystyle{aasjournal}



\end{document}